\documentclass[12pt]{amsart}
\usepackage{amsmath,amssymb,amsfonts,amsthm,graphicx,color,fullpage}

\begin{document}
\title{Calculation of Topological Charge of Real Finite-Gap sine-Gordon solutions using Theta-functional formulae}
\author{P. G. Grinevich and K. V. Kaipa}
\address{L.D. Landau Institute for Theoretical Physics, Russian Academy of Sciences}
\email{pgg@landau.ac.ru}
\address{Department of Mathematics, University of Maryland, College Park}
\email{kaipa@math.umd.edu}
\maketitle
\newcommand\bZ{\mathbb Z}
\newcommand\bR{\mathbb R}
\newcommand\bC{\mathbb C}
\newcommand\beq{\begin{equation}}
\newcommand\eeq{\end{equation}}

\newtheorem{lemma}{Lemma}
\newtheorem{theorem}{Theorem}
\newtheorem{definition}{Definition}
\newcounter{one}
\setcounter{one}{1}
\newcounter{six}
\setcounter{six}{6}
\newcounter{nineteen}
\setcounter{nineteen}{19}
\renewcommand{\Re}{\mathop{\mathrm{Re}}}
\renewcommand{\Im}{\mathop{\mathrm{Im}}}

\setlength\topmargin{-0.3in}

\noindent The most basic characteristic of real solutions $u(x,t)$ of the sine-Gordon equation $u_{tt}-u_{xx}+\sin u=0$ which are quasiperiodic in 
the $x$-variable is the {\it density of topological charge} defined as:\[ \bar n =\lim\limits_{T\rightarrow\infty}\frac{u(x+T,t)-u(x,t)} {2\pi T} \]
The real finite-gap solutions $u(x,t)$ are expressed in terms of the Riemann theta function of non-singular hyperelliptic 
curves $\Gamma: \mu^2 = \lambda \prod_{i=1}^{2g} (\lambda - E_i)$ and a generic positive divisor $D$ of degree $g$ on $\Gamma$, where the spectral 
data 
$(\Gamma, D)$ must satisfy some reality conditions. The problem of calculating $\bar n$ in terms of the spectral data was first 
studied in \cite{DubNov} and \cite{Nov}. A solution to this problem was obtained in \cite{GN} avoiding the use of the $\theta$-functional 
formula for $u(x,t)$. As pointed out by S.P. Novikov, if the $\theta$-functional form of solutions is to be considered as an effective
one, then it should be possible to calculate $\bar n$ directly from the $\theta$-functional formulae. 
We achieve this goal in the present note, using a new {\it multiscale} or {\it elliptic limit}  of real finite-gap sine-Gordon solutions.
The authors would like to express their gratitude to Professor Novikov for attracting their attention 
to this problem and stimulating discussions. \\

\noindent The reality condition on $\Gamma$ is that $\{E_1, \cdots, E_{2g}\} = \{ \overline{E_1}, \cdots, 
\overline{E_{2g}}\}$ for all $i$, and 
$E_i < 0$ if $E_i \in \bR$. The reality 
condition on the divisor found by Cherednik in \cite{Cher} is: $D+ \tau D -0 -\infty  = \mathcal{K}$ 
where $\tau(\lambda, \mu) = (\bar{\lambda}, \bar{\mu})$ and $\mathcal{K}$ is the canonical class. Such a divisor will be 
called  {\it admissible}. In order to write the $\theta$-functional formula for $u(x,t)$ we
need some notation. Let $a_i, b_j$ for $1 \leq i, j \leq g$ be a symplectic basis of cycles on $\Gamma$ and $\vec{\omega} = (\omega_1, \cdots,\omega_g)$ 
holomorphic differentials satisfying  $\int_{a_j} \omega_i =\delta_{i j}$. The Riemann matrix of $\Gamma$ is the matrix defined by $B_{ij}  = \int_{b_j}
\omega_i$. The Abel-Jacobi map $A: \Gamma \to J(\Gamma)$ is defined by $P \mapsto \int_{\infty}^P \vec{\omega}$, where 
$J(\Gamma)=\bC^g/\{\bZ^g + B \bZ^g\}$ is the Jacobian variety of $\Gamma$. Let $K \in J(\Gamma)$ be 
the associated vector of Riemann constants. Let $A(0) \equiv \epsilon'/2 + B \epsilon/2$, for some $\epsilon, \epsilon' \in \bZ^g$. We define  
$U=(\frac{\vec{\omega}}{d \sqrt{\lambda^{-1}}})(\infty)$ and $V=(\frac{\vec{\omega}}{d \sqrt{\lambda}})(0)$. The $\theta$-functional formula for 
$e^{i u(x,t)}$ can be written as follows (see \cite{KK}, \cite{DN}):
\beq 
e^{i u(x,t)} = C_1 \frac{ \theta( A(0)+ z(x,t))\, \theta( -A(0)+ z(x,t)) }{ \theta^2(z(x,t)) }
\eeq
where $z(x,t) =  -A(D)+ i x (V-U)/4 - i t (U+V)/4 - K$, $C_1^2 =  \exp( \pi  i \, \epsilon^t B \epsilon /2 )$, and 
\[ \theta(z|B) = \sum_{ n \in \bZ^g} \exp( \pi i \,n^t B n ) \, \exp(2 \pi i \,n^t z) \quad \mbox{ for } z \in \bC^g \]
In order to compute $\bar{n}$ for real solutions, we choose a special basis of cycles $a_i, b_j$  (figure (1) with parameter $k=1$) as 
suggested in 
\cite{GN}. Here $E_1, \cdots,E_{2m}$ are the non-zero real branch points and $E_{2j} = \overline{E_{2j-1}}$ for $ m+1 \leq j \leq g$. This basis 
satisfies $\tau a_i = - a_i$ for all $i$, $\tau b_i = b_i$ for $1 \leq i \leq m$ and $\tau b_i = b_i + a_i$ for $m+1 \leq i \leq g$, 
therefore $ \Re(B) = -1/2  \bigl( \begin{smallmatrix}   0 & 0\\   0 & I_{g-m} \end{smallmatrix} \bigr) $
where $I_{g-m}$ is the identity matrix of size $g-m$. (From here on, all vectors $v$ written as $(v_1, v_2)^t$ are understood to be split into 
blocks of length $m$ and $g-m$). We have $A(0) =  \epsilon'/2 + B \epsilon/2$ with $\epsilon' = (0,1)^t$ and $\epsilon = (1,0)^t$. 
The reality condition on the divisor becomes 
\beq \begin{split}  
A(D) &=  x + B \, { 1/2 - s/4 \choose  1/2}, \quad   s={ {s_1\atop s_2} \choose  {\cdot \atop \,s_m}} \\
{\mbox{ for some}} \quad x \in & \bR^g/ \bZ^g \quad \mbox{and} \quad {\mbox{ for some}} \;\;\; s_k \in \{-1,1\}
\end{split} \eeq
Thus the set $\{ z \in J(\Gamma) \, | \, z = -A(D)-K \quad \mathrm{ with } \; D \;\mathrm{ admissible } \}$ has $2^m$ components characterized by the 
symbol  $\vec{s} = (s_1, s_2, \cdots,s_m)$, each of which is a real $g$-dimensional torus denoted by $T_{\vec{s}}$. The vectors $i \,U, i \,V$ are 
purely 
real and hence $z(x,t) \in T_{\vec{s}}$. It is a consequence of the Cherednik reality condition that the $T_{\vec{s}}$ are disjoint from $ \Theta
\cup (\Theta -A(0))$ where $\Theta = \{ z \in  J(\Gamma) \, |\, \theta(z) = 0 \}$. Therefore the real solutions $u(x,t)$ given by equation (1) are 
non-singular. We also define functions $u_j(T)$ for $ T \in [0,1]$ by the formula: 
\[ e^{i u_j(T)} = C_1 \frac{ \theta( A(0)+ z_j(T))\, \theta( -A(0)+ z_j(T)) }{ \theta^2(z_j(T)) }, \qquad z_j(T) =  -A(D) - T e_j - K \]
where $e_j$ is the $j$-th standard basis vector of $\bC^g$, and therefore $z_j(T)$ represents the $j$-th basic cycle of 
$T_{\vec{s}}$. The topological charge density $\bar{n}$ is given by the formula (see \cite{GN}):
\[ \bar{n}= \sum_{j=1}^g (i \,U_j-i \,V_j) \, n_j /4, \quad \mathrm{ where } \quad n_j = \frac{1}{2 \pi i} \int_{T=0}^{T=1} d \log e^{i u_j(T)} \]
It is convenient to replace $\epsilon = (1,0)^t$ in the expression  $A(0) =\epsilon'/2 + B \epsilon/2$ with $\tilde{\epsilon}$ defined by   
$\tilde{\epsilon}_j = (-1)^j s_j$ for $ 1\leq j \leq m$ and $\tilde{\epsilon}_j = 0$ for $j > m$. Using the transformation rule
$\theta(z+N+BM) = \theta(z) \, \exp ( - 2 \pi i \, M^t z - \pi i \,  M^t B M)$ we get
\beq n_j = -\tilde{\epsilon}_j +  \frac{2}{2 \pi i} \int_{T=0}^1 d \log \left(  \frac{\theta( B \tilde{\epsilon}/2 + z_j(T) )}{ \theta(z_j(T)) 
}  \right) \eeq
where we have also used the fact that the cycles $ \epsilon'/2 + B \tilde{\epsilon}/2 + z_j(T) $ and $B \tilde{\epsilon}/2 + z_j(T) $ 
are homologous in the real torus $T_{-\vec{s}}$. The charges $n_j$ are easily calculated using formula (3) in two special cases: 
\begin{lemma}  In the case when $m=0$ (no real branch points) we have $n_j = 0$ for all $j$. \\
In the case $g=m=1$, we have $n_1 = s_1$. For later use, we also calculate $n_1$ from (3) if $K$ is taken to be $1/2$ instead of 
the correct value $(1+\tau)/2$, and $s_1$ is replaced by $-s_1$. In this case $n_1 = -s_1$. \end{lemma}
\noindent We reduce the calculation of the charges $n_j$ in the general case to the two special cases of Lemma 1. 
We consider the following  family of real nonsingular hyperelliptic curves $\Gamma(k)$ depending on the real parameter $k \in [1, \infty)$:
\[ \Gamma(k): \mu^2 = \lambda \prod_{i=1}^m ( \lambda - k^{i-1} E_{2i-1})  ( \lambda - k^{i-1} E_{2i}) \, \prod_{i=2m+1}^{2g}  ( \lambda - k^m E_i)\]
The basic cycles $a_i(k), b_j(k)$ on $\Gamma(k)$ are chosen as shown in figure (1).  The original curve 
$\Gamma$ coincides with $\Gamma(1)$. Let $\mathcal{C}_i$ for $1 \leq i \leq m$ be the real elliptic curves defined by 
$\mathcal{C}_i: y^2 = x (x- E_{2i-1}) ( x - E_{2i})$ and let $\mathcal{C}_{m+1}$ be the real hyperelliptic curve $\mathcal{C}_{m+1}: y^2 = x  
\prod_{i=2m+1}^{2g}  ( \lambda - E_i)$. Let $B(k)$ denote the Riemann matrix of the curve $\Gamma(k)$ with respect to the basic cycles  
$a_i(k), b_j(k)$. 
 \begin{figure}[htb]
\input{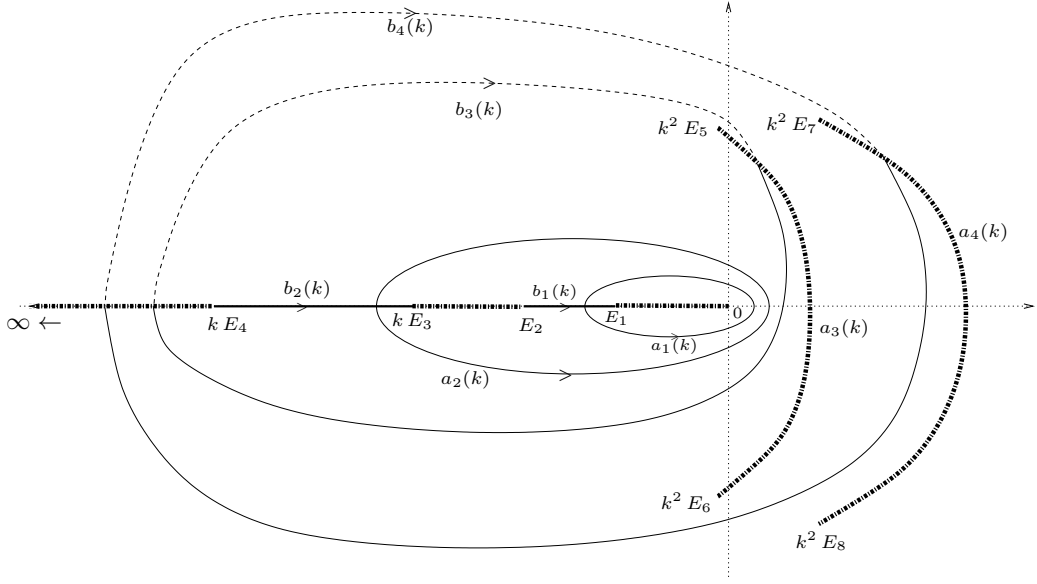} 
  \caption{ Basic cycles and cuts on $\Gamma(k)$ for $g=4, m=2$ }
\end{figure}

\begin{theorem} The limiting curve $\Gamma(\infty) = \lim_{k \to \infty} \Gamma(k)$ is a nodal curve 
with $m$ nodes when $m < g$, and $m-1$ nodes when $m=g$. The irreducible components of the normalization of $\Gamma(\infty)$ are 
$\{\mathcal{C}_i \,|\, 1 \leq i \leq m+1\}$ when $m <g$, and $\{\mathcal{C}_i \,|\, 1 \leq i \leq g\}$ when $m=g$. The limiting Riemann matrix 
$B(\infty) = \lim_{k \to \infty} B(k)$ is block diagonal $B(\infty) = 
$ diag$(B_1, B_2)$ where $B_1 = $ diag$(\tau_1, \cdots, \tau_m)$ for some purely imaginary $\tau_j$ in the complex upper half 
plane, 
and $\Re(B_2)=-1/2 \, I_{g-m}$.
\end{theorem} 
\noindent The charges $n_j$ being integers are constant during the deformation $k: 1 \to \infty$. Therefore $n_j$ can be calculated from formula 
(3) using the Riemann matrix $B(\infty)$.\\ \noindent Since $\theta((z_1,z_2)^t\, | $ diag$(B_1, B_2)) = \theta(z_1 |  B_1) \,\theta(z_2 | B_2)$, using 
Lemma 1 
together with the formula $ K = \sum_{i=1}^{g} A(E_{2i-1}) = \frac{1}{2} \,{ 1 \choose \nu_2} + \frac{1}{2} \, B { \nu_1 \choose 1 }$ where 
$(\nu_1,\nu_2)^t=(1,2,\cdots,g)^t$, we obtain the result: 
\beq
n_j = \begin{Bmatrix}  (-1)^{j-1} s_j  \qquad \mathrm{ if } \quad 1 \leq j \leq m \\ 0  \qquad \qquad \mathrm{ if } \quad j > m   \end{Bmatrix} 
\eeq

\noindent The details will appear in a future article. \\

\noindent {\it Remarks}
\begin{enumerate} \item In \cite{GN}, the admissible divisors were characterized by certain symbols $\{s_1', \cdots, s_m'\} \in \{ \pm 1\}^m$ 
defined as follows. Given an admissible divisor $D=\{ (\lambda_i, \mu_i) \,|\, 1 \leq j \leq g \}$ let $P(\lambda)$ be the unique polynomial of 
degree $g-1$ interpolating the $g$ points $(\lambda_i, \mu_i/\lambda_i)$. Then $P(\lambda)$ is real and $s_j'$ is defined to be 
the sign of $P(\lambda)$ over $[E_{2j}, E_{2j-1}]$. It was shown in \cite{GN} that the charges $n_j$ are equal to $(-1)^{j-1} s_j'$ for $j \leq m$ and $n_j=0$ 
for $j > m$. Comparing with formula (4), it follows that the symbols $s_j'$ and $s_j$ coincide for all $j$. 

\item The multiscale limit of the spectral curve constructed above was used only for a topological argument. The sine-Gordon solutions 
$u(x,t,k)$ associated with the spectral curve $\Gamma(k)$ (and admissible divisors $D(k)$) depend on the vectors $U(k)$ and $V(k)$ mentioned in the 
introduction. As $k \to \infty$, some component of $U(k)$ will diverge to $\infty$. Thus there is no limiting solution. However asymptotic expansion 
in the parameter $k$ of $u(x,t,k)$ involving elliptic (genus $1$) solutions can be written. This will be investigated in a future work.

\end{enumerate}


\begin{thebibliography}{99}



\bibitem{Cher} I.V. Cherednik, \textit{Reality conditions in ``finite-zone'' integration}, Sov. Phys. Dokl. \textbf{25} (1980), 450--452.


\bibitem{DN} B.A. Dubrovin, S.M. Natanzon, \textit{Real two-zone solutions of the sine-Gordon equation}, Funct. Anal. Appl. \textbf{16} (1982), 21--33.

\bibitem{DubNov} B.A. Dubrovin, S.P. Novikov,  \textit{Algebro-geometrical Poisson brackets for real finite-zone solutions of the Sine-Gordon 
equation and the nonlinear Schr\"odinger equation},  Sov. Math. Dokl.  \textbf{26} (1982), no.~3, 760--765.

\bibitem{GN} P.G. Grinevich, S.P. Novikov, \textit{Topological charge of the real periodic finite-gap sine-Gordon solutions}, Comm. Pure Appl. 
Math.  \textbf{56} (2003), no.~7  956--978.

\bibitem{KK} V.A. Kozel, V.P. Kotlyarov,  \textit{Almost periodic solutions of the equation } $u_{tt}-u_{xx}+\sin u=0$. (Russian), Dokl. Akad. Nauk 
Ukrain. SSR  Ser. A (1976),  no.~10 878--881.

\bibitem{Nov} S.P. Novikov, \textit{Algebrotopological approach to the reality problems. Real action variables in the theory of finite-zone 
solutions of the Sine-Gordon equation}, Zap. Nauchn. Sem. LOMI: Differential geometry, Lie groups and mechanics. \Roman{six}, 
\textbf{133}  (1984), 177--196.


\end{thebibliography}
\end{document}